\documentclass[pra,twocolumn,showpacs,nobibnotes,floatfix,superscriptaddress]{revtex4}

\usepackage{latexsym,amssymb,amsfonts,mathbbol,graphicx,color}
\usepackage{dcolumn}
\usepackage{bm}

\begin{document}

\title{Quantum Error Correction During 50 Gates} 
\author{Yaakov S. Weinstein}
\affiliation{Quantum Information Science Group, {\sc Mitre},
200 Forrestal Rd, Princeton, NJ 08540, USA}

\begin{abstract}
Fault tolerant protocol assumes the application of error correction after every quantum gate. However, correcting errors is costly in terms of time and number of qubits. Here we demonstrate that quantum error correction can be applied significantly less often with only a minimal loss of fidelity. This is done by simulating the implementation of 50 encoded, single-qubit, quantum gates within the [[7,1,3]] quantum error correction code in a noisy, non-equiprobable Pauli error environment with error correction being applied at different intervals. We find that applying error correction after every gate is rarely optimal and even applying error correction only once after all 50 gates, though not generally optimal, sacrifices only a slight amount of fidelity with the benefit of 50-fold saving of resources. In addition, we find that in cases where bit-flip errors are dominant, it is best not to apply error correction at all. 
\end{abstract}


\maketitle

Standard approaches to quantum fault tolerance (QFT), the computational framework that allows for successful quantum computation despite a finite probability of error in basic computational gates \cite{Preskill,ShorQFT,G,AGP}, assume the application of quantum error correction (QEC) \cite{book,ShorQEC,CSS} after every operation. QEC codes protect quantum information by storing some number of logical qubits in a subspace of a greater number of physical qubits thus forming the building blocks for QFT. However, the syndrome measurements needed to check for and correct errors are very expensive in terms of number of qubits required and implementation time. In this paper we demonstrate via numerical simulations that applying QEC after every operation is not necessary and, in general, not optimal. The simulations are done for single-logical-qubit operations on information encoded in the [[7,1,3]] QEC code \cite{Steane}. 

A guiding principle of QFT is to implement all protocols in such a way so as to ensure that information does not leave the encoded subspace (and become subject to errors). Only specialized gates can adhere to this principle. Nevertheless, for many QEC codes universal quantum computation can be performed within the QFT framework if the gate set is restricted to Clifford gates plus the $T$-gate, a single-qubit $\pi/4$ phase rotation. It is not {\it a priori} obvious how to implement general gates using such a restricted gate set. A method for implementing an arbitrary single-qubit rotation (within prescribed accuracy $\epsilon$) within these constraints was initially explored in \cite{SK1,SK2} and has recently become an area of intense investigation \cite{Svore1,KMM1,TMH,Svore2,KMM2,Selinger1,KMM3}. For Calderbank-Shor-Steane (CSS) codes, Clifford gates can be implemented bit-wise while the $T$-gate requires a specially prepared ancilla state and a series of controlled-NOT gates. Thus, the primary goal of these investigations has been to construct circuits within $\epsilon$ of a desired (arbitrary) rotation while limiting the number of resource-heavy $T$-gates. As an example, a $\sigma_z$ rotation by 0.1 can be implemented with accuracy better than $10^{-5}$ using 56 \cite{KMM3} $T$-gates, interspersed by at least as many single-qubit Clifford gates. QFT would suggest that QEC be applied after each one of the more than 100 gates needed to implement such a rotation requiring thousands of additional qubits and hundreds of time steps. Adhering to this is thus very resource intensive. 

Recently there have been a number of attempts to reduce the resource consumption of a quantum computation by carefully analyzing, simulating, and comparing protocols within the QFT  framework \cite{YSW,BHW,YSWTgate,NFB,YT}. Specifically, it was shown that QEC need not be applied after every gate and, in fact, should not be applied after every gate \cite{How}. Applying QEC less often will consume less resources, while still enabling successful quantum computation. This point was also made, though addressed in a different way, in Ref.~\cite{WIPK}. Here we numerically simulate the implementation of 50 logical gates on information encoded into the [[7,1,3]] QEC code applying QEC (via syndrome measurements and possible recovery operations) at different intervals and determining which scheme is best for different error probabilities. The simulations are explicit, the entire density matrix is calculated at every step.  

The [[7,1,3]] or Steane QEC code will correct an error on one physical qubit of a seven qubit system that encodes one qubit of quantum information. If errors occur on two (physical) qubits the code will be unable to restore the system to its proper state. By applying gates following the rules of QFT, we can ensure that the probability of an error occurring on two physical qubits remains of order $p^2$, where $p$ is the probability of a single qubit error per gate, no matter how many gates are applied. Thus, if $p$ is small enough one need only apply QEC at the end of the sequence. However, for long sequences of gates it is likely that  the coefficients in front of the higher order error terms will grow to an unacceptable level. QEC would then be needed more often. 

Of course, if QEC could be implemented perfectly, and we have unlimited resources available, it would be worthwhile to apply QEC as much as possible. In reality, QEC cannot be done perfectly and we are extremely concerned about resource consumption. Thus, we are left to ask, how often should QEC be applied?   

To address this we simulate 50 single-qubit gates on the [[7,1,3]] QEC code in a nonequiprobable Pauli operator error environment \cite{QCC} with non-correlated errors. As in \cite{AP}, this error model is a stochastic version of a biased noise model that can be formulated in terms of Hamiltonians coupling the system to an environment. Here, different error types arise with different arbitrary probabilities. Individual qubits undergo $\sigma_x^j$ errors with probability $p_x$, $\sigma_y^j$ errors with probability $p_y$, and $\sigma_z^j$ errors with probability $p_z$, where $\sigma_i^j$, $i = x,y,z$ are the Pauli spin operators on qubit $j$. We assume that only qubits taking part in a gate operation, initialization, or measurement will be subject to error while other qubits are perfectly stored. This idealized assumption is partially justified in that idle qubits may be less likely to undergo error than those involved in gates (see for example \cite{Svore}). 
   
We assume a single qubit state $|\psi\rangle=\cos\alpha|0\rangle+e^{i\beta}\sin\alpha|1\rangle$, perfectly encoded into the [[7,1,3]] error correction code. We then implement a series of gates, $U_{50}...U_2U_1$, in the nonequiprobable error environment leading to a final state, $\rho_f$, of the 7 qubits.  To determine the accuracy of the simulated implementations with perfectly applied gates, $\rho_i$, we utilize the state fidelity $F(\rho_i,\rho_f) = {\rm{Tr}}[\rho_i\rho_f]$. In addition we will find it useful to utilize the infidelity $I(\rho_i,\rho_f) = 1 - F(\rho_i,\rho_f)$. 

Our choice of gate sequence stems from the above noted work on the implementation of arbitrary single qubit gates with gates from the set Clifford plus $T$. We define the composite gates $A = HPT$ and $B = HT$ and simulate the implementation of the 50 gates:
\begin{equation}
U = ABBBAAAABBABABABBBAA.
\end{equation}
We then formulate 7 different error correction application schemes: applying QEC after every gate (50 QEC applications), after every composite gate $A$ and $B$ (20 applications), after every other composite gate (10 applications), after every 5 composite gates (4 applications), after each half of the sequence $U$ (2 applications), only after the entire sequence (1 application), and not at all. Each scheme is simulated for error environments of different values of $p_x$, $p_y$ and $p_z$. For the initial state we use the basis state $|0\rangle$. Other tested initial states and gate sequences give similar results.   

Implementing a Clifford gate, $C$, on the [[7,1,3]] QEC code requires implementing $C^{\dag}$ on each of the 7 qubits.
To implement a logical $T$-gate on a state encoded in the [[7,1,3]] QEC code requires constructing the ancilla state $|\Theta\rangle = \frac{1}{\sqrt{2}}(|0_L\rangle+e^{i\frac{\pi}{4}}|1_L\rangle)$, where $|0_L\rangle$ and $|1_L\rangle$ are the logical basis states on the [[7,1,3]] QEC code. Bit-wise CNOT gates are then applied between the state $|\Theta\rangle$ and the encoded state with the $|\Theta\rangle$ state qubits as control. Measurement of zero on the encoded state projects the encoded state with the application of a $T$-gate onto the qubits that had made up the $|\Theta\rangle$ state. Our simulations are done in a fault tolerant fashion following \cite{YSWTgate}.

\begin{figure}
\includegraphics[width=8.5cm]{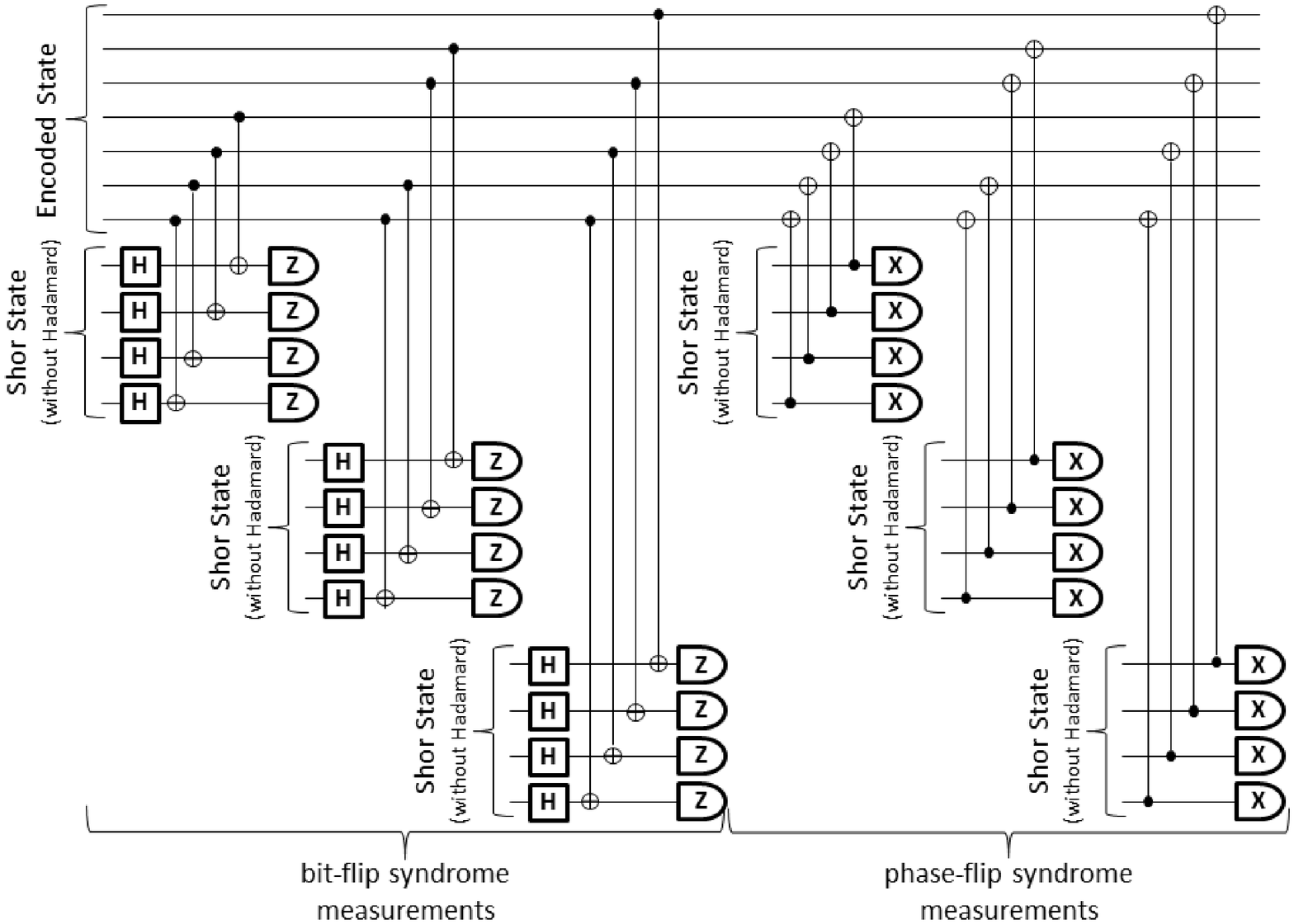}
\caption{Circuit for syndrome measurements for the [[7,1,3]] QEC code. Syndrome measurement is done in a fault tolerant fashion using four-qubit Shor states, GHZ states with a Hadamard applied to each qubit. The Shor states themselves are constructed in the nonequiprobable Pauli operator error environment. In addition, the set of bit-flip and phase-flip syndromes is repeated twice. }
\label{QEC}
\end{figure}

\begin{figure}
\includegraphics[width=8.5cm]{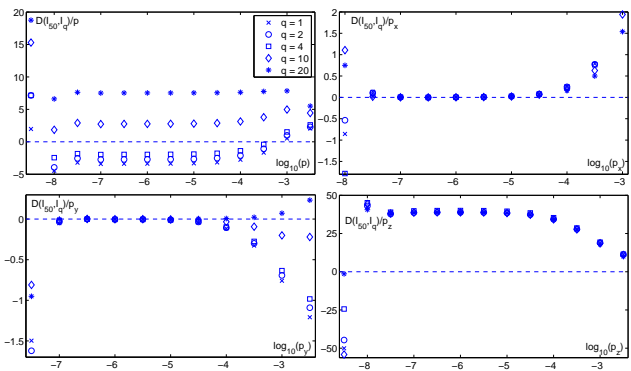}
\caption{Fractional change of logarithmic infidelity of $\rho_f$ divided by error probability, $p_i$, for 50 gates applied to initial state $|0\rangle$ with $q = 1$ ($\times$), 2 ($\bigcirc$), 4 ($\square$), 10 ($\diamond$), and 20 ($*$). The top left figure is for a depolarizing error environment ($p = p_x = p_y = p_z$) and the  remaining figures are for nonequiprobable error environments where $p_i$ is dominant and $p_j = p_k = 10^{-10}$: top right is for $i = x$, bottom right for $i = y$, and bottom right for $i = z$. }
\label{Every}
\end{figure}

Our simulation results are depicted in Tables \ref{Depol} and \ref{PErr} and Fig.~\ref{Every}. The two tables show results of each QEC application scheme for different error environments: Table \ref{Depol} for depolarization, $p = p_x = p_y = p_z$, and Table \ref{PErr} for error models where one error probability, $p_i$, is dominant and the other two remain constant $p_j = p_k = 10^{-10}$. In each Table the top line displays the output state infidelity when QEC is applied after every gate (50 times), $I_{50}$. Lower lines show the fractional change, $D$, in the infidelity upon using other QEC application schemes where:
\begin{equation}
D(I_{50},I_{q}) = \frac{I_{50}-I_{q}}{I_{50}}
\end{equation}   
and $q = 20, 10, 4, 2, 1, 0$. Note that a positive fractional change means that the infidelity is lower when using less QEC and thus the fidelity is higher. In other words, positive fractional change means a higher fidelity when using less QEC. Negative fractional change means the fidelity is higher when applying QEC after every gate. We quickly note, however, that even if applying QEC after every gate gives the highest fidelity, this does not mean it is the optimal choice of QEC application scheme. If the fractional change, $D(I_{50},I_1)$ is small one may achieve an almost optimal fidelity while saving a factor of up to 50 in time and number of qubits, perhaps a worthwhile tradeoff.   

Figure \ref{Every} extends the results shown in the tables by plotting $D(I_{50},I_q)/p$ as a function of error probability for the values of $q$ mentioned above and the four different error environments from the tables. The plots demonstrate which schemes are better (above zero) or worse (below zero) than applying QEC after every gate and by how much.  

Looking first at the infidelities when QEC is applied after every gate we find the following. In a depolarizing environment the infidelity increases steadily by an order of magnitude for every order of magnitude increase in error probability up to $p = 10^{-4}$, after which the increase is slightly faster. The same linear behavior is seen for the bit-flip error environment for $10^{-9} \le p_x \le 10^{-3}$, and for the phase-flip and $\sigma_y$ error environments for $10^{-8} \le p_y,p_z \le 10^{-4}$. Not surprisingly, the depolarizing environment most quickly decreases the fidelity. The other three environments, however, are not equal. The phase-flip and $\sigma_y$ dominated environments decrease the fidelity to about an equal extent, but the bit-flip environment is significantly more damaging. This inequality in fidelity decrease between the error environments was analyzed in \cite{WB}. 

The tables and figure demonstrate which QEC schemes achieve the highest fidelity for a large range of error probabilities. In a depolarizing environment for extremely low $p$ applying QEC after every gate will give the best fidelity (not shown). This is to be expected since the fidelity cost of applying QEC at these low error probabilities is minimal. For $p \simeq 10^{-9}$, applying QEC after every gate gives the lowest fidelity except for the case when no QEC is applied. As $p$ increases we must balance errors that have occurred during the implementation of the gates with errors that may arise from the error correction itself. We find that QEC is best applied often, but not too often. Two QEC schemes $q = 10, 20$ will give a higher fidelity than QEC after every gate. For $p > 10^{-3}$ applying QEC after every gate will give the lowest fidelity presumably because the cost of applying QEC outways the gain in correcting errors that arise during gate implementation. Thus, in the depolarizing environment, the best scheme for the range $10^{-9} \le p \le 10^{-2}$ is $q = 20$. The worst scheme generally is to apply error correction only once after all 50 gates. The worst case of this is when $p = 10^{-3.5}$ which, nevertheless, gives a fidelity only $1\times 10^{-5}$ lower than if QEC is applied after every gate. For every $10^{.5}$ decrease in $p$ the fidelity difference between the schemes of applying QEC after every gate and applying QEC only once will decrease by an order of magnitude such that at $p = 10^{-8}$ the difference in fidelity is only $2\times 10^{-14}$. Such a small change in fidelity may not warrant 50 times the time and number of qubits that would be necessary to apply QEC after every gate.

\begingroup
\begin{table*}
\caption{Second line: infidelity of final state after 50 noisy gates with noisy QEC applied after each as a function of depolarization strength $p = p_x = p_y = p_z$. Lower lines: fractional increase or decrease of infidelity for different QEC application schemes compared to the case of QEC after every gate.}
\begin{tabular}{|c|c|c|c|c|}
\hline 
QEC applications & $p = 10^{-6}$ & $p = 10^{-5}$ & $p = 10^{-4}$ & $p = 10^{-3}$ \\\hline\hline
50 & $4.50\times 10^{-5}$ & $4.54\times 10^{-4}$ & $4.90\times 10^{-3}$ & $8.27\times 10^{-2}$ \\\hline\hline
20 & $7.54\times 10^{-6}$ & $7.55\times 10^{-5}$ & $7.62\times 10^{-4}$ & $7.85\times 10^{-3}$ \\\hline
10  & $2.76\times 10^{-6}$ & $2.80\times 10^{-5}$ & $3.13\times 10^{-4}$ & $4.97\times 10^{-3}$\\\hline
4 & $-1.94\times 10^{-6}$ & $-1.89\times 10^{-5}$ & $-1.39\times 10^{-4}$ & $1.52\times 10^{-3}$\\\hline
2 & $-2.71\times 10^{-6}$ & $-2.65\times 10^{-5}$ & $-2.12\times 10^{-4}$ & $9.88\times 10^{-4}$\\\hline
1  & $-3.38\times 10^{-6}$ & $-3.31\times 10^{-5}$ & $-2.76\times 10^{-4}$ & $5.12\times 10^{-4}$\\\hline
0 & $-1.02$ & $-1.01$ & $-.941$ & $-.544$ \\\hline  
\end{tabular}
\label{Depol}
\end{table*}
\endgroup

\begingroup
\begin{table*}
\caption{Second line: one minus the fidelity of final state after 50 noisy gates with noisy QEC applied after each gate as a function of $p_i$ with $p_j = p_k = 10^{-10}$. Lower lines: percent increase or decrease of one minus the fidelity of different QEC application schemes compared to applying QEC after every gate as a function of $p_i$.}
\begin{tabular}{|c|c|c|c||c|c|c||c|c|c|}
\hline 
q & $p_x = 10^{-7}$ & $p_x = 10^{-5}$ & $p_x = 10^{-3}$ & $p_y = 10^{-7}$ & $p_y = 10^{-5}$ & $p_y = 10^{-3}$ & $p_z = 10^{-7}$ & $p_z = 10^{-5}$ & $p_z = 10^{-3}$ \\\hline\hline
50 & $3.1\times 10^{-6}$ & $3.1\times 10^{-4}$ & $3.2\times 10^{-2}$ & $7.0\times 10^{-7}$ & $7.0\times 10^{-5}$ & $1.1\times 10^{-2}$ & $7.0\times 10^{-7}$ & $7.1\times 10^{-5}$ & $1.9\times 10^{-2}$ \\\hline\hline
20 & $2.9\times 10^{-10}$ & $1.6\times 10^{-7}$ & $1.5\times 10^{-3}$ & $-1.6\times 10^{-10}$ & $6.7\times 10^{-9}$ & $7.2\times 10^{-5}$ & $3.9\times 10^{-6}$ & $3.8\times 10^{-4}$ & $1.8\times 10^{-2}$ \\\hline
10 & $1.4\times 10^{-10}$ & $2.0\times 10^{-7}$ & $1.9\times 10^{-3}$ & $-2.2\times 10^{-9}$ & $-3.7\times 10^{-8}$ & $-2.0\times 10^{-4}$ & $3.9\times 10^{-6}$ & $3.8\times 10^{-4}$ & $1.9\times 10^{-2}$ \\\hline
4  & $4.7\times 10^{-10}$ & $2.4\times 10^{-7}$ & $2.4\times 10^{-3}$ & $-2.1\times 10^{-9}$ & $-1.1\times 10^{-7}$ & $-6.3\times 10^{-4}$ & $4.0\times 10^{-6}$ & $4.0\times 10^{-4}$ & $1.9\times 10^{-2}$ \\\hline
2  & $3.9\times 10^{-10}$ & $2.4\times 10^{-7}$ & $2.4\times 10^{-3}$ & $-3.5\times 10^{-9}$ & $-1.1\times 10^{-7}$ & $-6.9\times 10^{-4}$ & $3.9\times 10^{-6}$ & $3.8\times 10^{-4}$ & $1.9\times 10^{-2}$ \\\hline
1  & $4.3\times 10^{-10}$ & $2.5\times 10^{-7}$ & $2.4\times 10^{-3}$ & $-5.0\times 10^{-9}$ & $-1.2\times 10^{-7}$ & $-7.6\times 10^{-4}$ & $3.8\times 10^{-6}$ & $3.8\times 10^{-4}$ & $1.9\times 10^{-2}$ \\\hline
0 & $9.5\times 10^{-2}$ & $9.6\times 10^{-2}$ & $8.1\times 10^{-2}$ & $-2.0$ & $-2.0$ & $-1.3$ & $-5.0$ & $-4.9$ & $-1.8$ \\\hline
\end{tabular}
\label{PErr}
\end{table*}
\endgroup

When the error model is asymmetric, $p_j = p_k = 10^{-10} < p_i$,  we see widely varying results depending on the degree of asymmetry and which errors are dominant. When $\sigma_y$ errors are dominant applying QEC after each gate is always optimal in output state fidelity and, in general, less QEC applications lead to lower fidelites. The only exception is that the $q = 20$ scheme provides better fidelity for $p_y \ge 10^{-5.5}$. 
When $\sigma_z$ errors are dominant, applying QEC after every gate gives the worst fidelity of any scheme (besides not applying QEC at all). All other schemes are about equal with the best fidelity achieved for $q = 4$. 
When $\sigma_x$ errors are dominant not applying QEC at all leads to the highest fidelity by far. For the other schemes, when $p_x \ge 10^{-5.5}$ the best fidelity is achieved by applying QEC just once. For lower values of $p_x$ the scheme giving the best fidelity is dependent on the exact value of $p_x$. The reason that no QEC is optimal for this error model is because the QEC code plus syndrome measurement approach used here are extremely sensitive to $\sigma_x$ errors than other types of errors \cite{WB}. Thus, the [[7,1,3]] QEC code with Shor state ancilla is not an appropriate error corretion approach for this error environment.  


In conclusion, our study calls into question the assumption that one must apply quantum error correction after every logical gate. Our simulations demonstrate that applying QEC after every logical gate will maximize the output state fidelity for only a limited set of error environments. Moreso, even in an error correction scheme where QEC after every gate does maximize the fidelity, the difference in fidelity between it and a scheme where QEC applied only once after 50 gates is minimal. It may be far more effective to choose the latter scheme so as to enjoy the 50-fold savings in time and number of qubits. 

The simulations reported here were done on a logical qubit of the [[7,1,3]] QEC code using Shor state ancilla for syndrome measurement. Using this error correction approach, we have shown that in a depolarization environment it is best not to apply QEC after every gate but instead after every composite $A$ and $B$ gate. When the errors are asymmetric the optimal choice of how many times to apply QEC will depend on which error is dominant and the size of the asymmetry. In cases where bit-flip errors dominate, we have shown that not applying QEC at all leads to the highest output state fidelity. This suggests that one should explore an alternate syndrome measurement strategy or QEC code. 

There are many choices to make in crafting a proper quantum error correction approach including an error correction code, syndrome extraction method, and encoding method. Here we have explored the question of how often to apply QEC, weighing the goal of achieving the highest output state fidelity with the additional cost of resources needed to apply QEC more often. Further work will explore other error correction codes and strategies with the goal of tailoring an optimal error correction approach to a given error environment.   

I would like to thank G. Gilbert for insightful comments. This research is supported under MITRE Innovation Program. 

\end{document}